\documentclass[usenatbib,a4paper]{mn2e}

\usepackage{graphicx}
\usepackage{natbib}
\newif\ifAMStwofonts
\usepackage{dcolumn}
\usepackage{bm}
\usepackage{amsmath}    
\usepackage{amsfonts}
\usepackage{amssymb}
\newcommand{\etal}{{\em et al.}}

\def\beqra{\begin{eqnarray}}
\def\eeqra{\end{eqnarray}}
\def\beq{\begin{equation}}
\def\eeq{\end{equation}}

\def\fr{\frac}
\def\pr{\prime}

\def \prrd {Phys.\ Rev.\ D\ }

\def\apj{Astrophys. J.\ }

\def\apjl{Astrophys. J. Lett.\ }
\def\apjs{Astrophys. J. Suppl.\ }
\def\mnras{Mon. Not. R. Astron. Soc.\ }

\def\physrep{Phys. Rept.\ }

\newcommand{\bi}{\begin{itemize}}
\newcommand{\ei}{\end{itemize}}
\newcommand{\be}{\begin{equation}}
\newcommand{\ee}{\end{equation}}
\newcommand{\bea}{\begin{eqnarray}}
\newcommand{\eea}{\end{eqnarray}}

\newcommand{\bfn}{\hat{\bf n}}

\newcommand{ \ie }{{\it i.e.}}

\def\spose#1{\hbox to 0pt{#1\hss}}
\def\approxgt{\mathrel{\spose{\lower 3pt\hbox{$\sim$}}
        \raise 2.0pt\hbox{$>$}}}

\renewcommand{\textsf}[1]{{\small #1}}

\newcommand\eq{\begin{equation}}
\newcommand\en{\end{equation}}

\def\edth{\;\raise1.0pt\hbox{$'$}\hskip-6pt\partial\;}
\def\baredth{\;\overline{\raise1.0pt\hbox{$'$}\hskip-6pt
\partial}\;}
\def\bi#1{\hbox{\boldmath{$#1$}}}
\def\gsim{\raise2.90pt\hbox{$\scriptstyle
>$} \hspace{-6.4pt}
\lower.5pt\hbox{$\scriptscriptstyle
\sim$}\; }
\def\lsim{\raise2.90pt\hbox{$\scriptstyle
<$} \hspace{-6pt}\lower.5pt\hbox{$\scriptscriptstyle\sim$}\; }

\title[Full-sky maps for gravitational lensing of the CMB]
{Full-sky maps for gravitational lensing of the CMB}
\author[]{%
Carmelita Carbone$^{1,2}$\footnotemark[1],
Volker Springel$^{3}$\footnotemark[2],
Carlo Baccigalupi$^{1}$\footnotemark[3],
Matthias Bartelmann$^{4}$\footnotemark[4],
\newauthor
Sabino Matarrese$^{5}$\footnotemark[5]
\\
\\ 
$^{1}$ SISSA/ISAS, Astrophysics Sector, Via Beirut 4, I-34014, Trieste, Italy and \\
INFN, Sezione di Trieste, Via Valerio, 2, 34127, Trieste, Italy \\
$^{2}$ Institut de Ci\`encies de l'Espai, CSIC/IEEC, Campus UAB, F. de Ci\`encies, Torre C5 par-2,  Barcelona 08193, Spain \\
$^{3}$ Max-Planck-Institute for Astrophysics, Karl-Schwarzschild-Str. 1, D-85741
Garching, Germany\\
$^{4}$ Institut f$\ddot{\rm u}$r Theoretische Astrophysik,
Universit$\ddot{\rm a}$t Heidelberg, Tiergartenstrasse 15, D-69121,
Heidelberg, Germany \\
$^{5}$ Dipartimento di Fisica `Galileo Galilei', Universit\`a di Padova and \\ 
INFN, Sezione di Padova, Via Marzolo 8, I-35131 Padova, Italy \\
}

\begin{document}

\maketitle

\begin{abstract}
We use the large cosmological Millennium Simulation (MS) to
construct the first all-sky maps of the lensing potential and the
deflection angle, aiming at gravitational lensing of the CMB, with the
goal of properly including small-scale non-linearities and
non-Gaussianity. Exploiting the Born approximation, we implement a
map-making procedure based on direct ray-tracing through the
gravitational potential of the MS. We stack the simulation box in
redshift shells up to $z\sim 11$, producing continuous all-sky maps
with arcminute angular resolution. A randomization scheme avoids
repetition of structures along the line of sight and structures larger
than the MS box size are added to supply the missing contribution of
large-scale (LS) structures to the lensing signal.  The angular power
spectra of the projected lensing potential and the deflection-angle
modulus agree quite well with semi-analytic estimates on scales down
to a few arcminutes, while we find a slight excess of power on small
scales, which we interpret as being due to non-linear clustering in
the MS.  Our map-making procedure, combined with the LS adding
technique, is ideally suited for studying lensing of CMB anisotropies,
for analyzing cross-correlations with foreground structures, or other
secondary CMB anisotropies such as the Rees-Sciama effect.
\end{abstract}

\begin{keywords}
gravitational lensing, cosmic microwave background, cosmology
\end{keywords}

\section{Introduction}
\label{i}
\renewcommand{\thefootnote}{\fnsymbol{footnote}}
\footnotetext[1]{E-mail: carbone@ieec.uab.es}
\footnotetext[2]{E-mail: volker@MPA-Garching.MPG.DE}
\footnotetext[3]{E-mail: bacci@sissa.it}
\footnotetext[4]{E-mail: mbartelmann@ita.uni-heidelberg.de}
\footnotetext[5]{E-mail: sabino.matarrese@pd.infn.it}
\renewcommand{\thefootnote}{\arabic{footnote}}

The cosmic microwave background (CMB) is characterized both by primary
anisotropies, imprinted at the last scattering surface, and by
secondary anisotropies caused along the way to us by density
inhomogeneities and re-scatterings on electrons that are freed during
the epoch of reionization, and heated to high temperature when massive
structures virialize.  One of the interesting effects that can
generate secondary anisotropies is the weak gravitational lensing of
the CMB, which arises from the distortions induced in the geodesics of
CMB photons by gradients in the gravitational matter potential
\citep{Matthias_rew,Lewis06}.  Forthcoming CMB probes do have the
sensitivity and expected instrumental performance which may allow a
detection of the lensing distortions of the primary CMB anisotropies,
which would then also provide new insights and constraints on the
expansion history of the universe and on the process of cosmological
structure formation \citep{acquaviva_baccigalupi_2006,hu_etal_2006}.
However, accurate predictions for the expected anisotropies in total
intensity and polarization are clearly needed for analyzing this
future data, which demands for detailed simulated maps.
 
The increasing availability of high-resolution N-body simulations in
large periodic volumes makes it possible to directly simulate the CMB
distortions caused by weak lensing using realistic cosmological
structure formation calculations. This work represents a first step in
that direction.  Existing studies already give access to statistical
properties of the expected all-sky CMB lensing signal, such as the
two-point correlation function and the power spectrum of the lensing
potential and deflection angle, see e.g. \citep{Lewis05} and
references therein. This is based on `semi-analytic' calculations that
use approximate parameterizations of the non-linear evolution of the
matter power spectrum. On the other hand, up to now N-body numerical
simulations have been used to lens the CMB only on small patches of
the sky in order to exploit the practicality of the flat-sky
approximation, see e.g. \citep{Amblard et al 04} and references
therein.  However, our approach of propagating rays through the
forming dark matter structures gives access to the full statistics of
the signal, including non-linear and non-Gaussian
effects. Furthermore, it allows the accurate characterization of
correlations of CMB lensing distortions with the cosmic large-scale
structure, and with other foregrounds such as the Sunyaev-Zeldovich
and Rees-Sciama effects. Hopefully this will allow improvements in the
methods for separating the different contributions to CMB anisotropies
in the data, which would be of tremendous help to uncover all the
cosmological information in the forthcoming observations.
\begin{figure*}
\begin{center}
\resizebox{8cm}{!}{\includegraphics{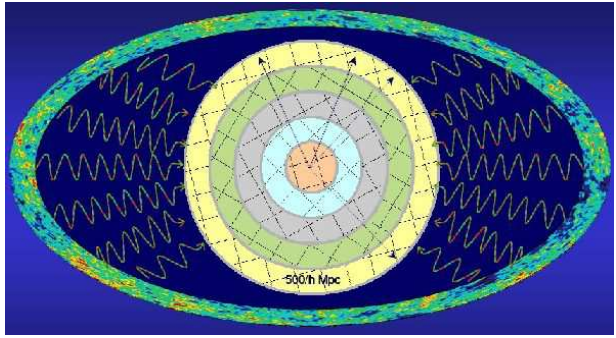}}
\caption{Sketch of the adopted stacking and randomization process. The
  passage of CMB photons through the dark matter distribution of the
  Universe is followed by stacking the gravitational potential boxes of
  the MS, which are $500\,h^{-1}{\rm Mpc}$ on a side (comoving). Shells
  of thickness $500\,h^{-1}{\rm Mpc}$ are filled with periodic replicas
  of the box. All boxes (squares) that fall into the same shell are
  randomized with the same coordinate transformation (rotation and
  translation), which, in turn, differs from shell to shell.}
\label{stacking}
\end{center}
\end{figure*}

From an experimental point of view, the improved precision of the CMB
observations, in particular that of the next generation
experiments\footnote{See \textsf{lambda.gsfc.nasa.gov} for a complete
list of operating and planned CMB experiments}, may in fact require an
accurate delensing methodology and a detailed lensing
reconstruction. CMB experiments targeting for instance the CMB
polarization, and in particular the curl component of the polarization
tensor, the so called $B$-modes from cosmological gravitational waves,
may greatly benefit from a precise knowledge of the lensing effects in
order to separate them from the primordial cosmological signal
\citep{seljak_hirata_2004}.  In particular, for a correct
interpretation of the data from the forthcoming Planck
satellite\footnote{www.rssd.esa.int/PLANCK}, it will be absolutely
essential to understand and model the CMB lensing, as the satellite
has the sensitivity and overall instrumental performance for measuring
the CMB lensing with good accuracy. We note that a first detection of
CMB lensing in data from the Wilkinson Microwave Anisotropy Probe
(WMAP\footnote{See \textsf{map.gsfc.nasa.gov}}) combined with
complementary data has already been claimed by
\citep{smith_etal_2007} and \citep{Seljak0108}.

In this study we introduce a new methodology for the construction of
all-sky lensing-potential and deflection-angle maps, based on a very
large cosmological simulation, the {\it Millennium} run
\citep{Springel2005}.  As a first step in the analysis of the
maps produced using the MS dark matter distribution, we have
determined the interval of angular scales on which these maps match
the semi-analytical expectations, since we expected a lack of lensing
power on large scales, due to the finite volume of the N-body
simulation.  To compensate for this effect, we have implemented a
method for adding large-scale power which allows to recover the
correct lensing signal on the scales outside this interval, i.e. on
scales larger than the MS box size.  At the other extreme, at the
smallest resolved scales, we are interested in the question whether
our maps show evidence for extra lensing power due to the accurate
representation of higher-order non-linear effects in our simulation
methodology. On these small scales, the impact of non-Gaussianities
from the mapping of non-linear lenses is expected to be largest.

This paper is organized as follows. In Section \ref{lmvtba}, we
briefly describe the basic aspects of lensing relevant to our work. In
Section \ref{mmpwtms}, we describe the N-body simulation and the
details of our map-making procedure. In Section~\ref{tslpada}, we
present the lensing-potential and deflection-angle maps, and study the
distribution of power in the angular domain. In Section~\ref{c} we
provide a summary and discussion.

\section{Lensed maps of the CMB via the Born approximation}
\label{lmvtba}

In what follows we will consider the \emph{small-angle scattering}
limit, \ie~the case where the \emph{change} in the comoving separation
of CMB light-rays, owing to the deflection caused by gravitational
lensing from matter inhomogeneities, is small compared to the comoving
separation of the \emph{undeflected} rays. In this case it is
sufficient to calculate all the relevant integrated quantities,
\ie~the so-called \emph{lensing-potential} and its angular gradient,
the \emph{deflection-angle}, along the undeflected rays. This
small-angle scattering limit corresponds to the so-called ``Born
approximation''.

We treat the CMB last scattering as an instantaneous process and
neglect reionization.  Adopting conformal time and comoving
coordinates in a flat geometry \citep{maber}, the integral for the
projected lensing-potential due to scalar perturbations with no
anisotropic stress reads
\begin{align}
\label{lensingpotential}
\Psi({\bf \hat{n}})\equiv 
-2\int_0^{r_*} \fr{r_*-r}{r_*r}\,\fr{\Phi(r{\bf\hat{n}};\eta_0-r )}{c^2}\,{\rm d}r\,,
\end{align}
while the corresponding deflection-angle integral is
\begin{align}
\label{deflection angle}
\boldsymbol{\alpha}({\bf \hat{n}})\equiv 
-2\int_0^{r_*}
\fr{r_*-r}{r_*r}\,\nabla_{\bf\hat{n}}\fr{\Phi(r{\bf\hat{n}};\eta_0-r )}{c^2}\, {\rm d}r\,,
\end{align}
where $r$ is the comoving distance, $r_*\simeq 10^4$ Mpc is its value
at the last-scattering surface, $\eta_0$ is the present conformal
time, $\Phi$ is the physical peculiar gravitational potential
generated by density perturbations, and $[1/r]\nabla_{\hat{\bf n}}$ is
the two dimensional (2D) transverse derivative with respect to the
line-of-sight pointing in the direction ${\hat{\bf
n}}\equiv(\vartheta,\varphi)$
\citep{Hu2000,Matthias_rew,Refregier,Lewis06}.

Actually, the lensing potential is formally divergent owing to the
$1/r$ term near $r=0$; nonetheless, this divergence affects the
lensing potential monopole only, which can be set to zero, since it
does not contribute to the deflection-angle. In this way the remaining
multipoles take a finite value and the lensing potential field is well
defined \citep{Lewis06}. Analytically, the full information about the
deflection angle is contained in the lensing potential, but
numerically the two equations (\ref{lensingpotential}) and
(\ref{deflection angle}) are generally not equivalent, and it will
typically be more accurate to solve the integral (\ref{deflection
angle}) directly to obtain the deflection angle instead of
finite-differencing the lensing potential.

If the gravitational potential $\Phi$ is Gaussian, the lensing potential
is Gaussian as well. However, the lensed CMB is non-Gaussian, as it is a
second order cosmological effect produced by cosmological perturbations
onto CMB anisotropies, yielding a finite correlation between different
scales and thus non-Gaussianity.  This is expected to be most important
on small scales, due to the non-linearity already present in the
underlying properties of lenses.\\

The most advanced approach developed so far for the construction of
all-sky lensed CMB maps \citep{Lewis05} employs a semi-analytical
modeling of the non-linear power spectrum \citep{Halofit}, and derives
from that the lensing potential and deflection angle templates
assuming Gaussianity. This approach is therefore accurate for what
concerns the two point correlation function of the lensing potential,
as long as the non-linear two-point power of the matter is modeled
correctly, but it ignores the influence of any statistics of higher
order, which is expected to become relevant on small scales, where the
non-linear power is most important.  The use of N-body simulations to
calculate the lensing has the advantage to possess a built-in
capability of accurately taking into account {\em all} the effects of
non-linear structure formation.  On the other hand, the use of N-body
simulations also faces limitations due to their limited mass and
spatial resolution, and from their finite volume, as we will discuss
later on in more detail.

For what concerns the line-of-sight integration in
Eqs.~\eqref{lensingpotential} and \eqref{deflection angle}, the
Born-approximation along the \emph{undeflected} photon path holds to
good accuracy and allows to obtain results which include the
non-linear physics. Even on small scales, in fact, this approximation
can be exploited in the small-angle scattering limit, \ie~for typical
deflections being of the order of arcminutes or less
\citep{Hirata:2003ka,Shapiro:2006em}. For example, a single cluster
typically gives deflection angles of a few arcminutes, while smaller
structures, such as galaxies, lead to arcsecond deflections.
Furthermore, it can be shown that the Born-approximation also holds in
`strong' lensing cases, provided that the deflection angles are
equally small. Finally, second order corrections to the Born
approximation (for instance a non-vanishing curl component) are
expected to be subdominant with respect to the non-linear structure
evolution effects on small scales \citep{Lewis06}.  For these reasons,
we argue that this approximation should be accurate enough for
calculating all-sky weak lensing maps of the CMB based on cosmological
N-body simulations.

\begin{figure*}
\begin{center}
\resizebox{16cm}{!}{\includegraphics{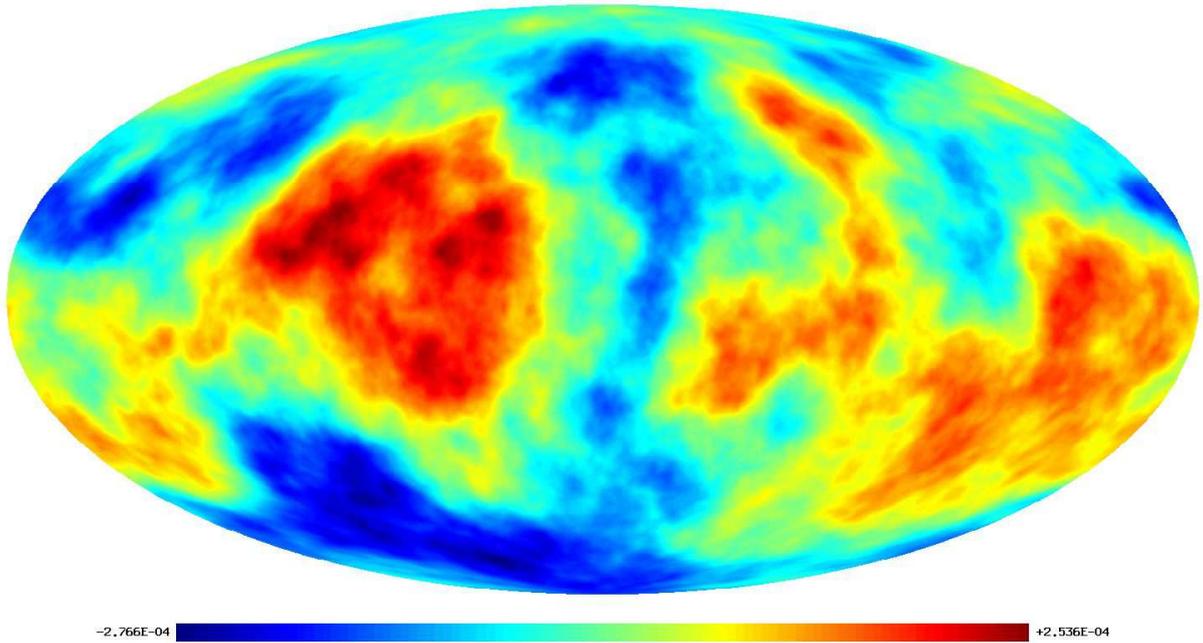}}
\footnotesize
\caption{The simulated all-sky map of the lensing potential computed
with the map-making procedure combined with the LS adding method as
described in the text.}
\label{projpot_map}
\end{center}
\end{figure*}
\begin{figure*}
\begin{center}
\resizebox{16cm}{!}{\includegraphics{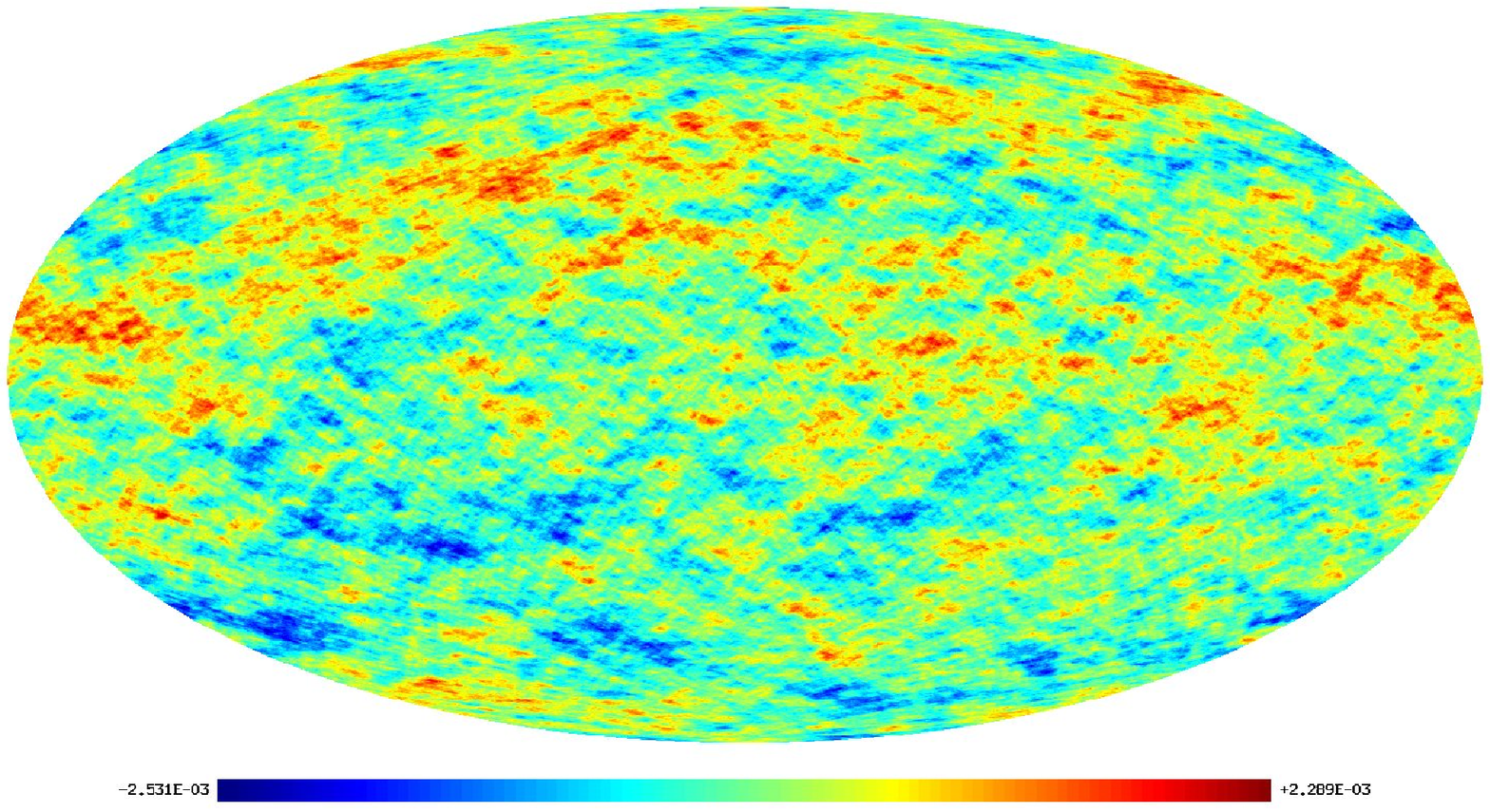}}
\resizebox{16cm}{!}{\includegraphics{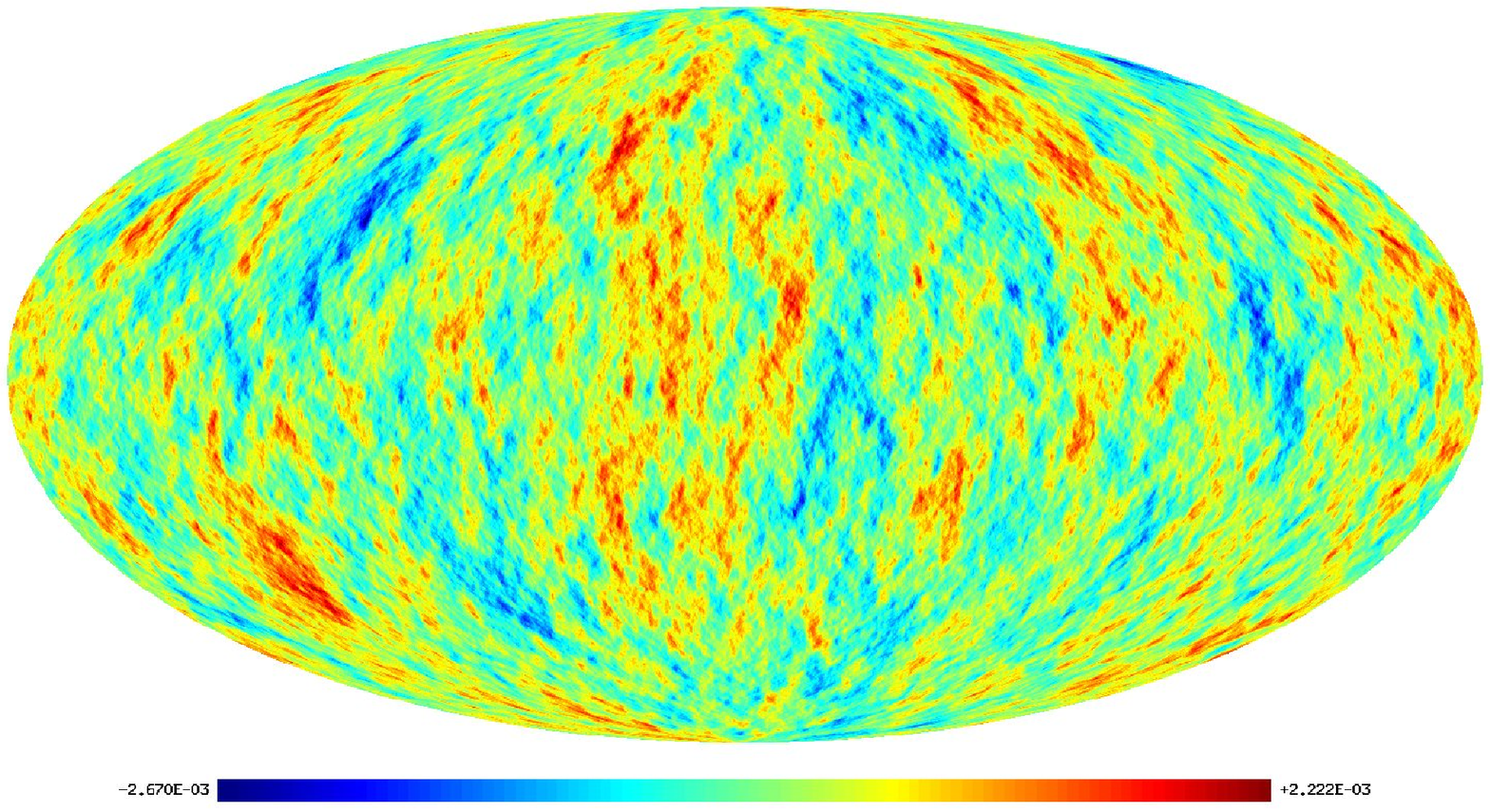}}
\footnotesize
\footnotesize
\caption{Simulated all-sky maps of the deflection-angle components
along the $\vartheta$ direction (top panel), and along the $\varphi$
direction (bottom panel), in radians.}
\label{angle1_map}
\end{center}
\end{figure*}
\begin{figure*}
\begin{center}
\resizebox{16cm}{!}{\includegraphics{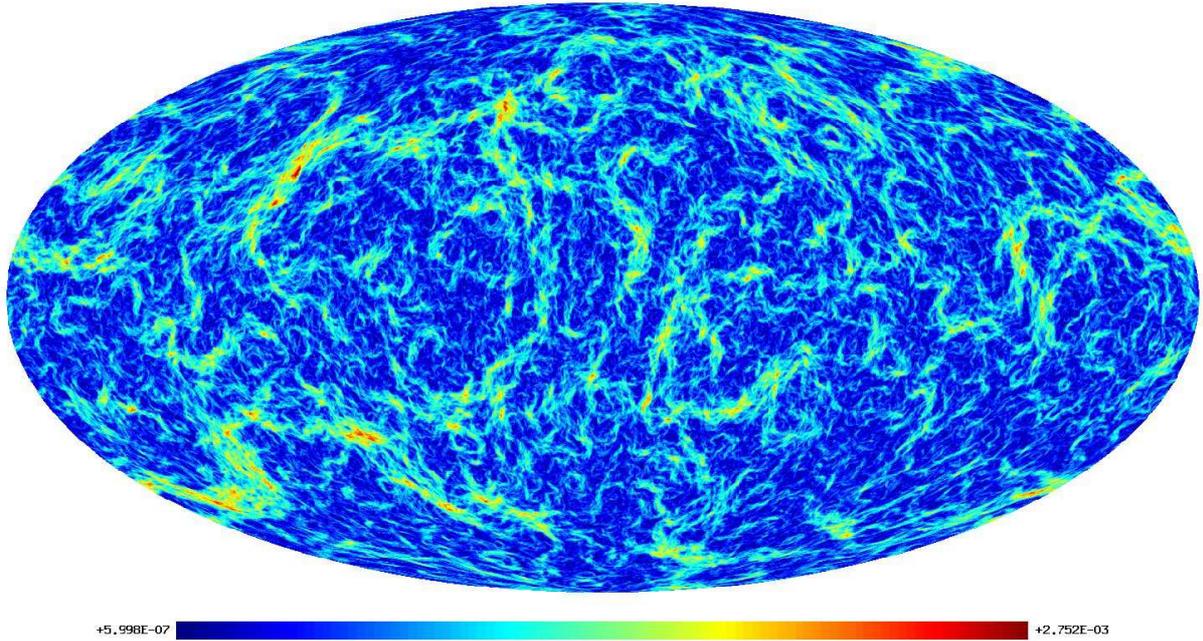}}
\footnotesize
\caption{{\em Top panel}: Simulated all-sky map of the
deflection-angle modulus (in radians), obtained with the map-making
procedure combined with the LS adding method as described in the
text.}
\label{anglemod_map}
\end{center}
\end{figure*}
\begin{figure*}
\begin{center}
\resizebox{11cm}{!}{\includegraphics{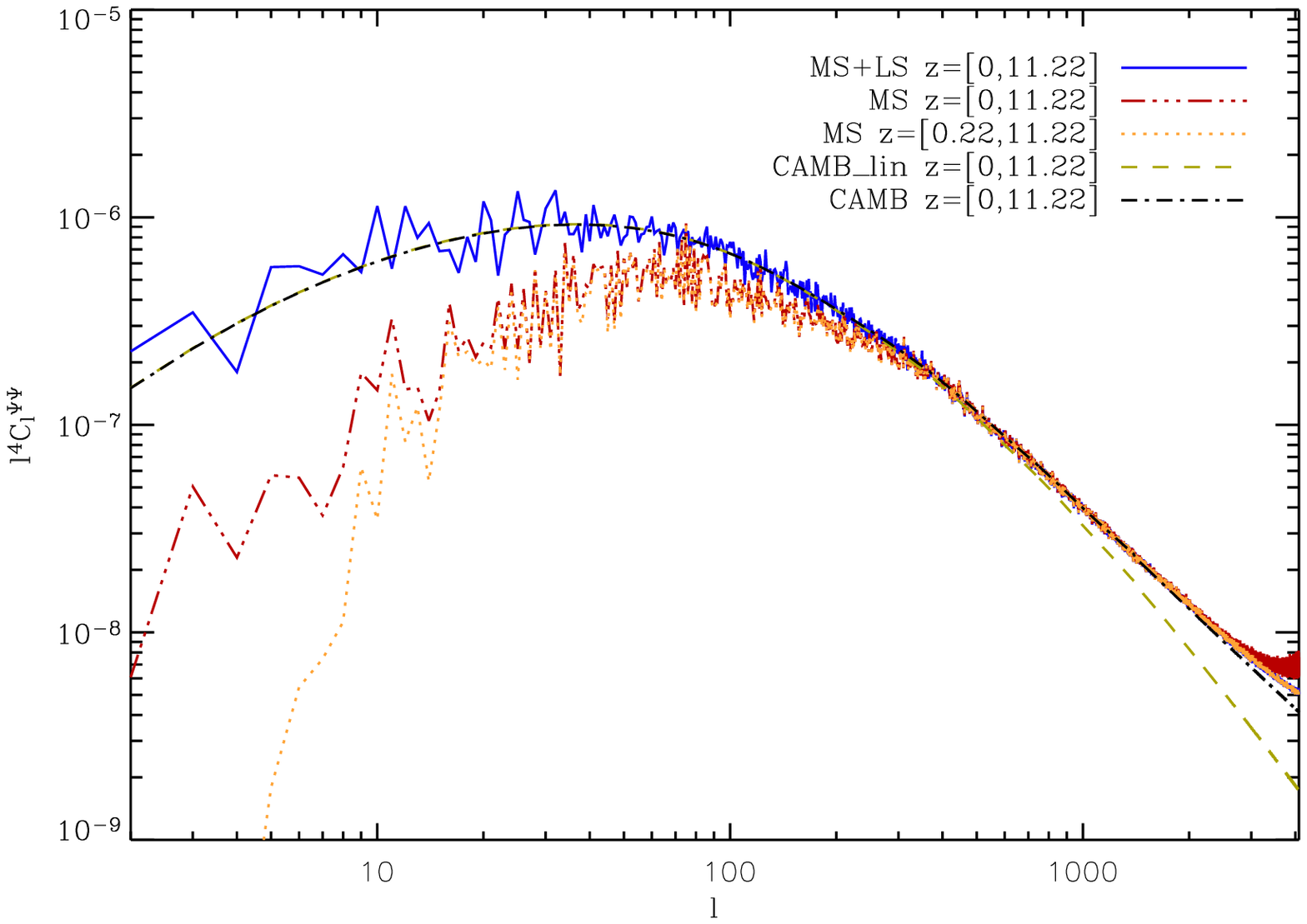}}
\resizebox{11cm}{!}{\includegraphics{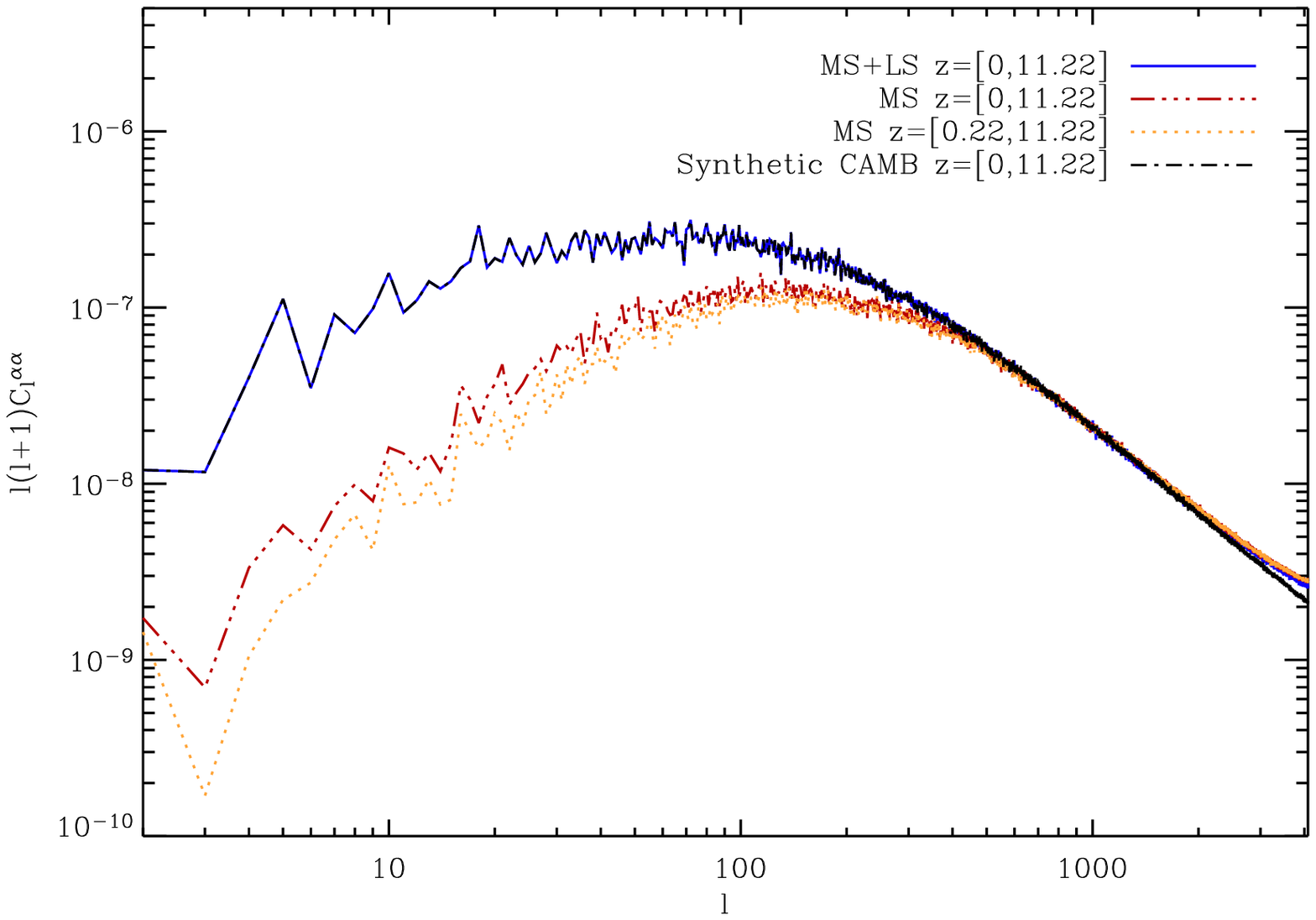}}
\footnotesize
\caption{{\em Top panel}: The power spectrum of the simulated lensing
potential map of Fig.~\ref{projpot_map} (blue solid line), compared
with the power spectrum of the lensing potential obtained with the
\textsf{CAMB} code (dashed-dotted black line), which also includes an
estimate of the non-linear contributions \citep{Halofit}. The red
dashed-3dotted and orange dotted lines differ only in the starting
redshift for the line-of-sight integration used in the
map-making. While the result shown in red begins at $z=0$, the orange
line gives the result for a start at $z=0.22$. Finally, the light-green
dashed line represents the linear lensing potential power spectrum from the
\textsf{CAMB} code.  {\em Bottom panel}: The power spectrum (in
radians squared) of the simulated deflection angle modulus map shown
in the upper panel of Fig.~\ref{anglemod_map} (blue solid line),
compared with the power spectrum (dashed-dotted black line) of the synthetic
deflection angle modulus map obtained as a Gaussian realization from
the \textsf{CAMB} lensing potential power spectrum.  As above, the red
dashed-3dotted and orange dotted lines differ only in the starting
redshift of the line-of-sight integration, as labelled. The red line
is for the full redshift interval, the orange one for a start at
$z=0.22$, as described in the text.}
\label{projpot_ps}
\end{center}
\end{figure*}

\section{Map-making procedure for the Millennium Simulation}
\label{mmpwtms}

The Millennium Simulation (MS) is a high-resolution N-body simulation
carried out by the Virgo Consortium \citep{Springel2005}. It uses $N=
2160^3\simeq 1.0078\times 10^{10}$ collisionless particles, with a
mass of $8.6\times 10^8\,h^{-1}{\rm M}_{\odot}$, to follow structure
formation from redshift $z=127$ to the present, in a cubic region
$500\,h^{-1}{\rm Mpc}$ on a side, and with periodic boundary
conditions. Here $h$ is the Hubble constant in units of $100\,{\rm
km\,s^{-1}Mpc^{-1}}$.  With ten times as many particles as the
previous largest computations of this kind
\citep{Colberg2000,Evrard2002,Wambsganss2004}, it features a
substantially improved spatial and time resolution within a large
cosmological volume.

The cosmological parameters of the MS are as follows. The ratio between
the total matter density and the critical one is $\Omega_{m}=0.25$, of
which $\Omega_{b}=0.045$ is in baryons, while the density of cold dark
matter (CDM) is given by $\Omega_{\rm CDM}=\Omega_{m}-\Omega_{b}$. The
spatial curvature is assumed to be zero, with the remaining cosmological
energy density made up by a cosmological constant,
$\Omega_{\Lambda}=0.75$. The Hubble constant is taken to be
$H_0=73\,{\rm km\,s^{-1} Mpc^{-1}}$. The primordial power spectrum of
density fluctuations in Fourier space is assumed to be a simple
scale-invariant power law of wavenumber, with spectral index
$n_s=1$. Its normalization is set by the rms fluctuations in spheres of
radius $8h^{-1}$ Mpc, $\sigma_8=0.9$, in the linearly extrapolated
density field at the present epoch.  The adopted parameter values are
consistent with a combined analysis of the 2dF Galaxy Redshift Survey
(2dfGRS) and the first year WMAP data \citep{Colless2001,spergel2003}.

Thanks to its large dynamic range, the MS has been able to determine
the non-linear matter power spectrum over a larger range of scales
than possible in earlier works \citep{Jenkins_etal_1998}. Almost five
orders of magnitude in wavenumber are covered
\citep{Springel2005}. This is a very important feature for
studies of CMB lensing, as we expect that this dynamic range, combined
with the method for adding large-scale structures described in the
next section, can be leveraged to obtain access to the full
non-Gaussian statistics of the lensing signal, limited only by the the
maximum angular resolution resulting from the gravitational softening
length and particle number of MS.  We stress again that the lensed
CMB is non-Gaussian even if the underlying lenses do possess a
Gaussian distribution. Moreover, the non-linear evolution of large
scale structures produces a degree of non-Gaussianity in the lenses
distribution which contributes to the non-Gaussian statistics of the
lensed CMB on small scales.  This non-Gaussian contribution can be
computed only via the use of N-body simulations which are able to
accurately describe the non-linear evolution of the lenses.  These
non-linearities are known to alter the lensed temperature power
spectrum of CMB anisotropies by about $\sim 0.2\%$ at $\ell\sim 2000$
and by $\sim 1\%$ or more on smaller scales. But, much more notably,
they introduce $\sim 10\%$ corrections to the B-mode polarization
power on {\it all the scales} \citep{Lewis05, Lewis06}.

Our map-making procedure is based on ray-tracing of the CMB photons in
the Born approximation through the three-dimensional (3D) field of the
peculiar gravitational potential. The latter is precomputed and stored
for each of the MS output times on a Cartesian grid with a mesh of
dimension $2560^3$ that covers the comoving simulation box of volume
$(500\,h^{-1}{\rm Mpc})^3$. The gravitational potential itself has been
calculated by first assigning the particles to the mesh with the
clouds-in-cells mass assignment scheme. The resulting density field has
then been Fourier transformed, multiplied with the Green's function of
the Poisson equation in Fourier space, and then transformed back to
obtain the potential. Also, a slight Gaussian smoothing on a scale $r_s$
equal to 1.25 times the mesh size has been applied in Fourier space in
order to eliminate residual anisotropies on the scale of the mesh, and a
deconvolution to filter out the clouds-in-cells mass assignment kernel
has been applied as well. The final potential field hence corresponds to
the density field of the MS (which contains structures down to the
gravitational softening length of $5\,h^{-1}{\rm kpc}$) smoothed on a
scale of $\simeq 200\,h^{-1}{\rm kpc}$.

In order to produce mock maps that cover the past light-cone over the
full sky, we stack the peculiar gravitational potential grids around the
observer (which is located at $z=0$), producing a volume which is large
enough to carry out the integration over all redshifts relevant for CMB
lensing.  For simplicity, we only integrate out to $z_*=11.22$ in this
study, which corresponds to a comoving distance of approximately
$r_*\sim 7236\,h^{-1}{\rm Mpc}$ with the present choice of cosmological
parameters.  Indeed, the lensing power from still higher redshifts than
this epoch is negligible for CMB lensing, as we will discuss in the next
section. But we note that our method could in principle be extended to
still higher redshifts, up to the starting redshift $z=127$ of the
simulation.

The above implies that the simulation volume needs to be repeated
roughly $14.5$ times along both the positive and negative directions of
the three principal Cartesian axes $x$, $y$, and $z$, with the origin at
the observer. However, the spacing of the time outputs of the MS
simulation is such that it corresponds to an average distance of
$140\,h^{-1}{\rm Mpc}$ (comoving) on the past light-cone. We fully
exploit this time resolution and use 53 outputs of the simulation along
our integration paths. In practice this means that the data
corresponding to a particular output time is utilized in a spherical
shell of average thickness $140\,h^{-1}{\rm Mpc}$ around the observer.

The need to repeat the simulation volume due to its finite size
immediately means that, without augmenting large-scale structures, the
maps will suffer from a deficit of lensing power on large angular
scales, due to the finite MS box size. More importantly, a scheme is
required to avoid the repetition of the same structures along the line
of sight. Previous studies that constructed simulated light-cone maps
for small patches of the sky typically simply randomized each of the
repeated boxes along the past lightcone by applying independent random
translations and reflections \citep[e.g.][]{Springel2001}. However, in
the present application this procedure would produce artefacts like
ripples in the simulated deflection-angle field, because the
gravitational field would become discontinuous at box boundaries,
leading to jumps in the deflection angle. It is therefore mandatory
that the simulated lensing potential of our all sky maps is everywhere
continuous on the sky, which requires that the 3D tessellation of the
peculiar gravitational potential is continuous transverse to every
line of sight.

Our solution is to divide up the volume out to $z_*$ into spherical
shells, each of thickness $500\,h^{-1}{\rm Mpc}$ comoving (obviously
the innermost shell is actually a sphere of comoving radius
$250\,h^{-1}{\rm Mpc}$, centered at the observer). All the simulation
boxes falling into the same shell are made to undergo the same,
coherent randomization process, \ie~they are all translated and
rotated with the same random vectors generating a homogeneous
coordinate transformation throughout the shell. But this randomization
changes from shell to shell.  Figure~\ref{stacking} shows a schematic
sketch of this stacking process.  For simplicity, the diagram does not
illustrate the additional shell structure stemming from the different
output times of the simulation. As discussed before, this simply means
that the underlying potential grid is updated on average ~3-4 times
with a different simulation output when integrating through one of the
rotated and translated $500\,h^{-1}{\rm Mpc}$ shells, but without
changing the coordinate transformation.  Notice that our stacking
procedure eliminates any preferred direction in the simulated all-sky
maps.

In order to define the gravitational potential at each point along a
ray in direction $\bfn$, we employ spatial tri-linear interpolation in
the gravitational potential grid. It is then easy to numerically
calculate the integral potential for each ray, based for example on a
simple trapezoidal formula, which we use in this study. Obtaining the
deflection angle could in principle be done by finite differencing a
calculated lensing potential map, either in real space or the harmonic
domain. However, the accuracy of this approach would depend critically
on the angular resolution of the map. Also, the sampling of the
gravitational potential in the direction transverse to the
line-of-sight varies greatly with the distance from the observer, so
in order to extract the maximum information from the simulation data
down to the smallest resolved scales in the potential field, we prefer
to directly integrate up the deflection angle vector along each light
ray in our map. For this purpose we first use a fourth-order
finite-differencing scheme to compute the local 3D grid of the
gradient of the gravitational potential, which is then again
tri-linearly interpolated to each integration point along a
line-of-sight. In this way, we calculate the deflection angle directly
via equation \eqref{deflection angle} along the paths of undeflected
light rays.

Finally, we need to select a pixelization of the sky with a set of
directions ${\bf\hat{n}}\equiv(\vartheta,\varphi)$.  We here follow
the standard approach introduced by the
\textsf{HEALPix}\footnote{healpix.jpl.nasa.gov} hierarchical
tessellation of the unit sphere \citep{Healpix}.

\section{Simulated maps of the lensing potential and deflection angle}
\label{tslpada}

In Figs.~\ref{projpot_map}, \ref{angle1_map}, and
\ref{anglemod_map}, we show full-sky maps of the lensing potential,
the deflection angle $\vartheta$/$\varphi$-components, and
the deflection angle modulus $|\boldsymbol\alpha|$, respectively,
obtained with the map-making technique described in the previous
section combined with a semi-analytic procedure (to be explained
below) augmenting the lensing power on scales beyond the MS box size.
These maps are generated with a \textsf{HEALPix} pixelization
parameter $N_{\rm side}=2048$, and have an angular resolution of $\sim
1.72^\prime$ \citep{Healpix}, with 50331648 pixels in total.

Several interesting features should be noted in these maps. The
distribution of the lensing potential, where the monopole and dipole
have been cut to simplify the visual inspection, appears to be
dominated by large features, which are probably simply arising from
the projection of the largest scale gravitational potential
fluctuations along the line-of-sight.  However, the strength of local
lensing distortions in the CMB cannot be directly inferred from the
map of the lensing potential, as for the lensing deflection only the
gradient of the potential is what really matters.

The maps showing the lensing deflection angle components have
interesting features as well. First of all, the signal in the two
components of the deflection angle appears to possess two
morphologically distinct regimes, characterized on one hand by a
diffuse background distribution, caused probably by the lines-of-sight
where no dominant structures are encountered, and on the other hand by
sharp features, caused probably by massive CDM structures which give
rise to the largest deflections in the line-of-sight integration
itself. The same features are evident in the map of the modulus of the
deflection-angle.

The mean value of $|\boldsymbol\alpha|$ in our simulated maps is
$2.36^{\pr}$, while its standard deviation is $1.25^{\pr}$.
The latter has to be compared with the corresponding 
value obtained via the angular differentiation 
of synthetic Gaussian maps produced with the lensing potential
power spectrum generated by the
publicly available Code for Anisotropies in the Microwave
Background (\textsf {CAMB}\footnote{See \textsf{camb.info}.}) 
using the MS cosmological parameters, as we explain in detail below.
We find only a $0.03\%$ difference 
for the rms of the $|\boldsymbol\alpha|$-maps
from MS and \textsf{CAMB}, when using the same maximum 
redshift of line-of-sight integration, 
i.e. $z_{\textrm{max}}=11.22$. On the other hand, 
if we set $z_{\textrm{max}}=1100$ in \textsf{CAMB},
we find that our estimate is $\sim 1.7\%$ 
smaller than the semi-analytic one, 
due to the missed contribution from sources beyond $z\sim 11$ in our 
map-making procedure. 
For comparison, we also evaluate semi-analytically the expected 
change in the standard deviation of $|\boldsymbol\alpha|$
when inserting in \textsf{CAMB} more recent estimates 
of the cosmological parameters \citep{WMAP5}. 
In this case the rms from MS is $\sim 6\%$ and 
$\sim 4.2\%$ greater than the semi-analytical prediction when 
in CAMB we set $z_{\textrm{max}}=11.22$ 
and $z_{\textrm{max}}=1100$, respectively.

The lensing potential and deflection angle maps of 
Figs.~\ref{projpot_map} and \ref{anglemod_map} have been
obtained combining the map-making procedure described in the previous
section with the method for adding large-scale power that we now
explain.\\ Firstly, we have measured the power spectra of the simulated
maps obtained from the MS scales only, \ie, using the routine
\textsf{ANAFAST} of the \textsf{HEALPix} package, we have
independently measured the power spectra of the lensing potential
($C_l^{\Psi\Psi}$) and deflection angle modulus ($C_l^{\alpha\alpha}$)
of the MS simulated maps, without exploiting the relations between the
lensing potential and the $\pm1$-spin components of the deflection angle, 
which hold in the spherical harmonic
domain \citep{Hu2000}. Secondly, using the MS cosmological parameters, we have
evaluated the semi-analytical power spectrum of the lensing potential
from \textsf{CAMB},
including the estimate of the contribution from non-linearity
\citep{Halofit} and stopping the line-of-sight integration 
redshift up to $z=11.22$.
Using the lensing potential power spectrum from
{\small CAMB}, we have then produced the corresponding synthetic map
(and its angular differentiation) obtained as a Gaussian realization
generated with the \textsf{HEALPix} code \textsf{SYNFAST}, in order to
produce the synthetic map of the deflection angle modulus from the
semi-analytic expectations of {\small CAMB}.  From this map we have
then extracted the power spectrum of the deflection angle modulus, and
after deconvolution from the \textsf{HEALPix} pixel window function, we
have compared it, together with the lensing potential power spectrum,
to the corresponding deconvolved MS power spectra.

\begin{figure}
\begin{center}
\resizebox{8cm}{!}{\includegraphics{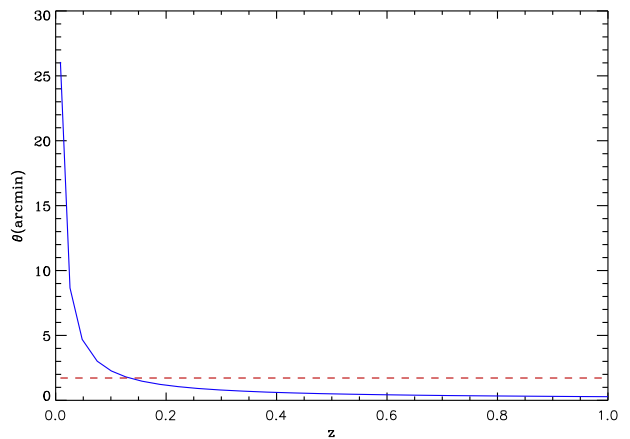}}
\footnotesize
\caption{Comparison between the angular resolution of $1.72^{\prime}$ of
our full-sky maps (red dashed line) and the redshift-dependent,
effective angular resolution (blue solid line) corresponding to the
intrinsic grid spacing ($\sim 200\,h^{-1}{\rm kpc}$) of the
three-dimension gravitational potential field constructed from the
Millennium Simulation.}
\label{resolution}
\end{center}
\end{figure}

The top panel of Fig.~\ref{projpot_ps} shows the primary result of
this comparison.  The black dashed-dotted line represents the semi-analytic
prediction of the lensing potential angular power spectrum obtained
from \textsf{CAMB} as discussed above. This has been compared with the
red dashed-3dotted line obtained with the map-making procedure previously
described, and which represents the result for the full integration
starting at $z=0$ and ending at $z=11.22$. In this case, a power
deficiency on large scales with respect to the semi-analytical
prediction is evident, and confined to a multipole range corresponding
to one degree or more in the sky.  The same for the orange dotted line
which gives the MS lensing potential power spectrum obtained from a
line-of-sight integration starting at a redshift of $z=0.22$ and
ending at $z=11.22$; comparing the two curves, a power decrease at low
$\ell$ is easily observable in the orange dotted line, with respect to
the red dashed-3dotted one, illustrating the influence of the lack of
comoving scales greater than $500\,h^{-1}{\rm Mpc}$ in the MS. As
expected, this effect is evident in the multipole range corresponding
to a few degrees or more, which is about the size of the MS box at the
redshift most relevant for CMB lensing, i.e. $z\simeq1$.  However,
towards larger $\ell$, the deficit of large-scale power quickly
decreases, and becomes negligible at scales $l \approxgt 350$.
Between these scales and $l\sim 2500$, there is quite good agreement
between the MS lensing power spectrum and the semi-analytic
prediction, but at $2500 \lesssim l \lesssim 4000$ the full MS signal
for the lensing potential actually slightly exceeds the semi-analytic
result. On this multipole range, the red dashed line is dominated by Poisson noise, 
but the slight excess of power is clearly observable from the orange dot line, in which
there is no contribution from the low redshift integration at $z \lesssim 0.2$.
We ascribe this power excess to the matter non-linearities
accurately reproduced from the Millennium Simulation. Finally, at
$l\sim 4000$ the MS signal is dominated by Poisson sampling noise from low-redshift
potential integration. In fact, at very low redshifts, 
the $1.72^{\prime}$ angular resolution of our map
is comparable and even smaller than the intrinsic angular resolution
corresponding to the spatial grid of the 3D gravitational potential
field we use. This is evident in Fig.~\ref{resolution}, where we
compare the map angular resolution of $1.72^{\prime}$ (red dashed
line) with the effective angular resolution corresponding to the
intrinsic grid spacing ($195\,h^{-1}{\rm kpc}$) of the 3D
gravitational potential field as function of redshift.  Because the
line-of-sight integral for the projected lensing potential involves a
$1/r$ weighting term, the resulting noise terms are unfavourably
amplified when the lensing potential is considered.

The comparison above has been used to evaluate the multipole range, $0
\lesssim l \lesssim 350$, not covered by the MS scales.  On this
interval we have applied the LS adding method: from the CAMB and MS
maps of the lensing potential, we have extracted the two corresponding
ensembles $\Psi_{lm}^{CAMB}$ and $\Psi_{lm}^{MS}$ of spherical
harmonic coefficients, respectively.  Since on low multipoles the
effects of the non-Gaussianity from the non-linear scales are
negligible and the $\Psi_{lm}$ are independent, we have generated a
joined ensemble of $\tilde{\Psi}_{lm}$, where
$\tilde{\Psi}_{lm}=\Psi_{lm}^{CAMB}$ for $0\leq l \leq 350$ and
$\tilde{\Psi}_{lm}=\Psi_{lm}^{MS}$ for $l > 350$.  Finally, we have
generated the synthetic maps of the lensing potential and deflection
angle as non-Gaussian constrained realizations, inserting the
$\tilde{\Psi}_{lm}$ as input in \textsf{SYNFAST}, as shown in
Figs.~\ref{projpot_map}-\ref{anglemod_map}.

These maps have the peculiarity of reproducing the non-linear and
non-Gaussian effects of the MS non-linear dark matter distribution at
multipoles $l > 350$, while at the same time including the
contribution from the large scales at $l \le 350$, where the lensing
potential follows mostly the linear trend as shown from the light-green
dot-dashed line in Fig.~\ref{projpot_ps}.  The blue solid curve in the
same Figure represents the resulting power spectrum of the lensing
potential map after the LS addition.

The bottom panel of Fig.~\ref{projpot_ps}, shows the corresponding
power spectra for the physically and numerically more meaningful
deflection angle.  Here we show a comparison of the power spectrum of
the deflection angle modulus ($C_l^{\alpha\alpha}$) measured for the
MS simulated maps, in the absence of LS supplying, with the
semi-analytic prediction constructed with \textsf{CAMB} and
\textsf{SYNFAST}, as explained above.  Again, we find a deficit of
power on large scales, and a reassuring agreement over about one order
of magnitude in $l$ on intermediate scales.  However, a slight excess
of power over the semi-analytic predictions is easily seen at $l
\approxgt 2500$. As previously mentioned, it can be attributed to the
non-linear evolution of the MS structures. Finally, the blue solid
line represents the power spectrum extracted from the deflection angle
modulus map of Fig.~\ref{anglemod_map}, after adding large-scale
structures.\\

\begin{figure}
\begin{center}
\resizebox{8cm}{!}{\includegraphics{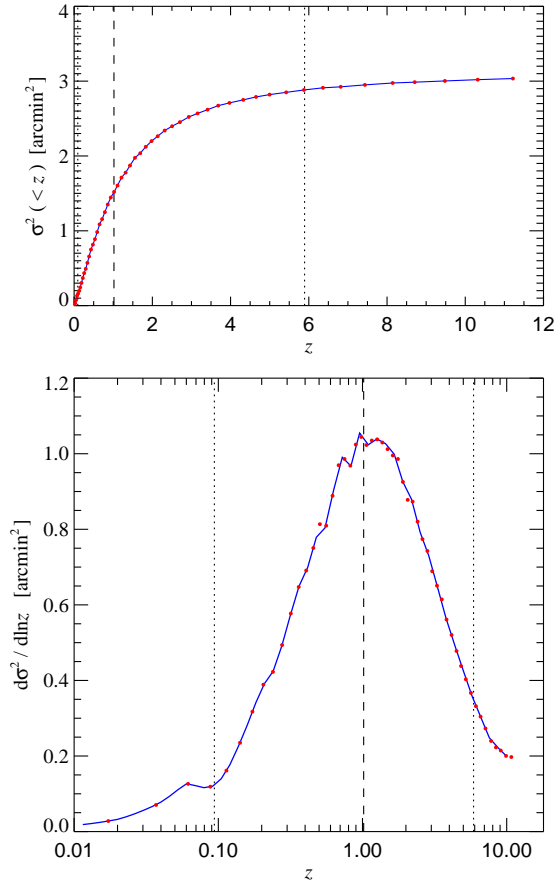}}\\
\footnotesize
\caption{The cumulative and differential variance of the deflection
  angle map as a function of redshift. The symbols mark the different
  output times of the Millennium Simulation.  The vertical dashed line
  gives the redshift that corresponds to the 50\% quartile of the
  total variance in our maps, which is approximately at $z\sim 1$. The
  dotted lines mark the 5\% and 95\% percentiles, indicating that 90\%
  of our signal in the deflection angle power spectrum is produced in
  the redshift range $z\sim 0.1$ to $z\sim 6.0$. Note however that we
  have lost a few percent of the total power due to our truncation of
  the integration at $z=11.22$. When included, this would slightly shift
  these precentiles to higher redshift.}
\label{standard_dev}
\end{center}
\end{figure}

Our map making procedure offers very good resolution at the most
important redshift for lensing of the CMB, $z\sim 1$ (see also
Fig.~\ref{standard_dev}), where the intrinsic angular resolution of
our potential grid is six times better than the angular resolution of
the full-sky map. We therefore think that this higher small-scale
power is a direct result of the more accurate representation of
non-linear structure formation in our map simulation methodology. In
fact, in our current maps we are still far from probing the most
non-linear scales accessible in principle with our simulation. Those
are a factor 40 smaller (namely $5\,h^{-1}{\rm kpc}$) than resolved by
the potential grid we have employed. However, using such a fine mesh
is currently impractical, and would lead to angular resolutions in
full-sky maps that are unaccessible even by the Planck
satellite. However, for a smaller solid-angle of the map, these scales
can be probed with a different ray-tracing technique
\citep{Hilbert2007}.\\

We note that the semi-analytic prediction for the power spectrum of
the deflection angle modulus has been evaluated as an angular gradient
in the harmonic domain of a synthetic lensing potential Gaussian map;
that is accurate since in this approach we work with Fourier modes
right from the start anyway.  From a numerical point of view, the
integral and derivative operators in Eq.~(\ref{deflection angle}) do
however not commute, even if they analytically do, in the sense that
finite differencing our measured projected potential will not
necessarily give the same result as numerically integrating the
deflection angle along each line of sight. The latter approach is more
accurate, expecially at very high resolution, and it has been used by
us in the comparison above since numerically integrating the
deflection angle along each line of sight allows to preserve the
contribution from the non-linear scales in a more efficient way than
simply operating in the harmonic domain. \\

Finally, we consider the distribution of the deflection angle power
along the line-of-sight.  In Fig.~\ref{standard_dev}, we show the
cumulative and differential variance of the deflection angle as a
function of redshift.  We see that the most important contributions to
the final signal stem from $z\sim 1$, i.e.~about half ways between the
last scattering surface and the observer, as expected.  This also
allows us to assess the relative error introduced by stopping the
integration at $z\simeq 11$, which is of the order of a few percent, 
as mentioned above.

\section{Conclusions}
\label{c}

We constructed the first all sky maps of the cosmic microwave
background (CMB) weak-lensing potential and deflection angle based on a
high-resolution cosmological N-body simulation, the Millennium Run
Simulation (MS). The lensing potential and deflection angle are
evaluated in the Born approximation by directly ray-tracing through a
three-dimensional, high-resolution mesh of the evolving peculiar
gravitational potential and its gradient. The time evolution is
approximated by 53 simulation outputs between redshift $z=0$ and
$z\simeq 11$, each used to cover a thin redshift interval
corresponding to a shell in the past light-cone around the observer.
To prevent artificial repetition of structures along the
line-of-sight, while at the same time avoiding discontinuities in the
force transverse to a line-of-sight, we tessellate shells of comoving
thickness corresponding to the size of the box ($500\,h^{-1}{\rm
Mpc}$) with periodic replicas which are coherently rotated and
translated within each shell by a random amount.  Moreover, in
order to include the contribution to the lensing signal from the
scales larger than the MS box size, we have implemented a method for
adding large-scale structure as described in the text.

Using the Hierarchical Equal Area Latitude Pixelization
(\textsf{HEALPIX}) package for obtaining a uniform sky-coverage, we
have constructed simulated CMB lensing maps with $\sim 5$ million
pixels and an angular resolution of $\sim 1.72^\prime$, based on
potential fields calculated on $2560^3$ meshes from the Millennium
simulation.  In the present study, we analyze the power spectrum of
the lensing potential and the deflection angle, and compare it with
predictions made by semi-analytic approaches. We note that our general
approach for map-making can be extended to other CMB foregrounds,
including the Integrated Sachs-Wolfe (ISW) and Rees-Sciama effects at
low redshifts, as well as estimates of the Sunyaev Zel'dovich (SZ)
effects, or of the X-ray background.  This will in particular allow
studies of the cross-correlation of the lensing of CMB temperature and
polarization with these effects, which will be the subject of a
forthcoming study.  In our approach we do not take into account
the contributions of the baryonic physics to the lensing effects on
the CMB. We expect in fact that these contributions could be
non-negligible only on the typical scales of cluster cores and below,
thus well above $l\sim 3000$.
Our comparison of the angular power spectrum of the
lensing-potential and the deflection-angle with semi-analytic
expectations reveals two different regimes in our results. First, for
multipoles up to $l\sim 2500$, our simulated maps produce a lensing
signal that matches the semi-analytic expectation.  Second, we find
evidence for a slight excess of power in our simulated maps on scales
corresponding to few arcminutes and less, which we attribute to the
accurate inclusion of non-linear power in the Millennium
simulation. It will be especially interesting to study the
non-Gaussianities in the signal we found and its implied consequences
for CMB observations.

The new method proposed here demonstrates that an all-sky mapping of CMB
lensing can be obtained based on modern high-resolution N-body
simulations. This opens the way towards a full and accurate
characterization of CMB lensing statistics, which is unaccessible beyond
the power spectrum with the existing semi-analytical techniques. This is
relevant in view of the forthcoming CMB probes, both as a way to detect,
extract and study the CMB lensing signal, which carries hints on the
early structure formation as well as the onset of cosmic acceleration,
and as a tool to distinguish CMB lensing from the Gaussian contribution
due to primordial gravitational fluctuations.

\section*{Acknowledgments}
We warmly thank L. Moscardini, A. Refregier, A. Stebbins, L. Verde and
Simon~D.~M.~White for helpful discussions and precious suggestions, and
M. Roncarelli for useful considerations.  Some of the results in this
paper have been derived using the Hierarchical Equal Area Latitude
Pixelization of the sphere (HEALPix, G\'orski, Hivon and Wandelt 1999).


\end{document}